# Nonlinear optical signal generation mediated by a plasmonic azimuthally chirped grating

Parijat Barman,[†,‡] Abhik Chakraborty,[†,‡] Denis Akimov,[†,‡] Ankit Kumar Singh,[‡] Tobias Meyer-Zedler,[†,‡] Xiaofei Wu,[‡] Carsten Ronning,[⊥] Michael Schmitt,[†,‡] Jürgen Popp,[†,‡] Jer-Shing Huang[†,‡,§,#]

[†]Institute of Physical Chemistry and Abbe Center of Photonics, Friedrich-Schiller-Universität Jena, Helmholtzweg 4, D-07743 Jena, Germany

[‡]Leibniz Institute of Photonic Technology, Albert-Einstein Str. 9, 07745 Jena, Germany

[⊥]Institut für Festkörperphysik, Friedrich-Schiller-Universität Jena, Max-Wien-Platz 1, 07743 Jena, Germany

[§]Research Center for Applied Sciences, Academia Sinica, 128 Sec. 2, Academia Road, Nankang District, Taipei 11529, Taiwan

[#]Department of Electrophysics, National Yang Ming Chiao Tung University, Hsinchu 30010, Taiwan

**ABSTRACT:** The deployment of plasmonic nanostructures to enhance nonlinear signal generation requires effective far-to-near field coupling and phase matching for frequency conversion. While the latter can be easily achieved at plasmonic hotspots, the former is an antenna problem that requires dedicated structural design and optimization. Plasmonic gratings are a simple but effective platform for nonlinear signal generation since they provide a well-defined momentum for photon-plasmon coupling and local hotspots for frequency conversion. In this work, a plasmonic azimuthally chirped grating (ACG), which provides spatially resolved broadband momentum for photon-plasmon coupling, was exploited to investigate the plasmonic enhancement effect in two nonlinear optical processes, namely two-photon photoluminescence (TPPL) and second-harmonic generation (SHG). The spatial distributions of the nonlinear signals were determined experimentally by hyperspectral mapping with ultrashort pulsed excitation. The experimental spatial distributions of nonlinear signals agree very well with the analytical prediction based solely on photon-plasmon coupling with the momentum of the ACG, revealing the "antenna" function of the grating in plasmonic nonlinear signal generation. This work highlights the importance of the antenna effect of the gratings for nonlinear signal generation and provides insight into the enhancement mechanism of plasmonic gratings in addition to local hotspot engineering.

Nonlinear effects in light-matter interaction arise from the higher-order corrections to the susceptibility of matter.[1] The nonlinear dependence of the induced polarization on the incident electric field has been used extensively for frequency conversion, which leads to a broad range of applications,[2,3] including telecommunications,[4] optical signal processing,[5] all-optical switching,[6] optical microscopy,[7] optical tomography,[8] and biosensing.[9] To facilitate such nonlinear processes, rationally designed nanostructures have been exploited for the engineering of both near and far optical fields related to nonlinear optical processes.[10-13] The success of nanostructure-assisted nonlinear signal generation lies in well-engineered overlapping and coupling of electromagnetic fields at multiple frequencies in both far- and near-field regimes. In general, the energy of far-field excitation at the fundamental frequency needs to be effectively concentrated in the near-field region, where the hotspot promotes the frequency conversion, and the locally generated nonlinear signals, possibly at multiple frequencies, need to be sent to the far-field region with high efficiency. Although the process is more complex than conventional nonlinear signal generation in bulk nonlinear crystals without nanostructures, it is the use of nanostructures that enables design-based enhancement of nonlinear signal generation.[14,15]

Plasmonic gratings are simple but effective nanostructures for enhancing nonlinear signal generation.[16] There are three key steps for plasmonic grating-based nonlinear signal generation (Fig. 1a). In the first step, the plasmonic grating should be designed such that it functions as an effective "receiving antenna" for the far-field excitation to generate near-field hotspots. This is typically done by choosing the right grating periodicity to provide the correct momentum for photon-plasmon coupling at the desired frequencies. In the second step, the hotspots should provide extreme spatial confinement and intensity enhancement of the optical near field at the input and output frequencies for the nonlinear optical processes. The spatial distribution of the hotspots at the input and output frequencies should overlap in space to provide maximal efficiency.[11] Since the hotspots are extremely confined in space, they provide broad spatial frequency bandwidth to facilitate the phase matching required for the frequency conversion.[17-26] In the third step, the grating should serve as a "transmitting antenna" to provide out-coupling channels for the locally generated nonlinear energy to enter the far field for detection.

Taking SHG as an example, the whole process can be illustrated in the energy-momentum space, as shown in Figure 1b. In the first step, a far-field (FF) photon at the fundamental frequency is converted into the near-field (NF) surface plasmon at the same frequency, i.e., FF ($\omega$)→NF($\omega$) by borrowing the lattice momentum of the grating, $mG$. Here, $m$ is the resonance order of the grating, and G is the grating momentum at the fundamental resonance. The function of the plasmonic grating in this step is similar to that of a receiving optical antenna. In the second step, the near-field hotspots provide broadband momentum for the phase matching, and the frequency conversion is achieved in the near-field regime, i.e., NF($\omega$)→ NF($2\omega$). In the third step, the grating functions as a transmitting optical antenna to send the near-field energy at the second harmonic (SH) frequency to the far field, i.e., NF($2\omega$)→FF ($2\omega$). While the near-field nonlinear signal generation in the second step is promoted by plasmonic hotspots, the coupling processes in the first and the third steps require the grating to be an effective optical antenna to mediate the far and near optical fields. Therefore, the grating structures must be carefully designed to achieve the necessary coupling efficiency and near-field hotspots.[12, 19-26]

Among various nonlinear signals, TPPL and SHG are two nonlinear processes that commonly occur in plasmonic nanostructures. While both nonlinear signals exhibit a quadratic dependence on the excitation power, the underlying mechanisms are quite different. TPPL is a third-order nonlinear process involving the sequential absorption of two photons as input and the emission of a blue-shifted single photon as output.[17, 27-29] The first photon absorption results in an intra-band transition of an electron in the conductive $sp$-band. The second photon sequentially brings about an inter-band transition of a second electron from the d-band to fill the hole created by the first photon in the $sp$-band (Figure 1a).[17, 27, 28] For the intra-band transition in the first step, momentum conservation can only be fulfilled with the broadband momentum provided by the plasmonic hotspots.[17, 18, 27, 28] The recombination of the electron from the first transition with the hole in the $d$-band occurs through a radiative relaxation into broadband TPPL with shorter wavelengths than the excitation. On the other hand, SHG is a second-order nonlinear process marked by output at the SH of the fundamental excitation frequency.[2] SHG requires the potential profile of the oscillating electrons to be symmetry broken.[2] It is inherently weak and constrained by the phase-matching conditions. Therefore, the SHG of materials with central symmetry, such as metals, is usually weak under typical excitation by plane waves in the far field.

However, SHG can be significantly enhanced by the hotspots of resonant plasmonic nanostructures due to field enhancement and localization. As explained previously, the field enhancement compensates for the weak nonlinearity of high-order harmonic generation, and the extreme field localization provides broadband momentum necessary for the required phase-matching condition.[30] In order to promote nonlinear signal generation, plasmonic nanostructures need to be carefully designed.

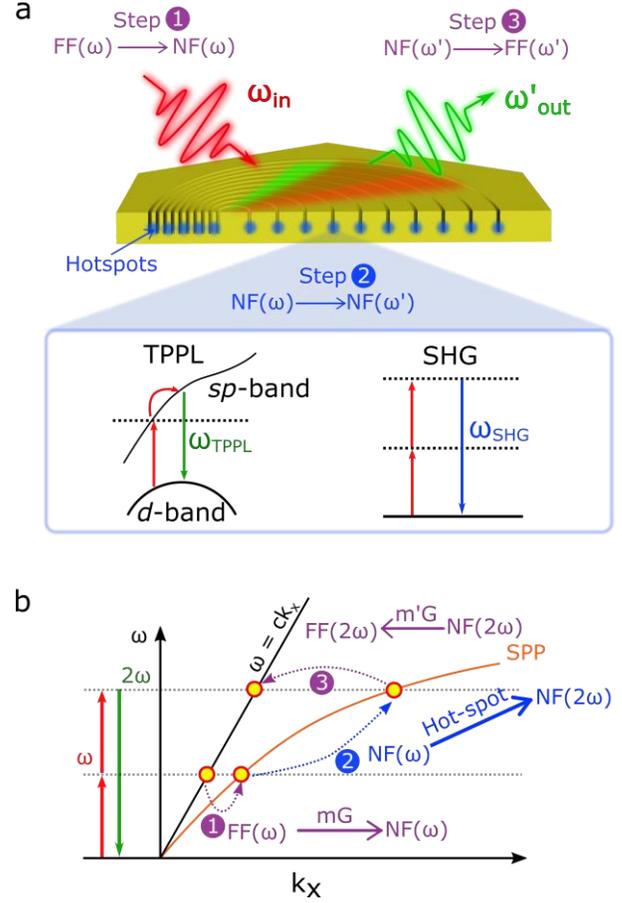

**Figure 1.** (a) Schematic illustration of the transitions involved in TPPL and SHG. (b) Energy-momentum plot illustrating the role of the plasmonic grating in different steps of plasmonic SHG. In steps (1) and (3), the grating functions as an optical antenna to mediate the far field (FF) and near field (NF) by providing the grating momentum (G) at specific resonance order ($m$) for photon-plasmon coupling at the desired frequencies. In step (2), the hotspots in the grating groove support the phase matching and facilitate the near-field frequency conversion, i.e., NF($\omega$)→NF($2\omega$).

In this work, the enhancement effects of a plasmonic ACG in TPPL and SHG were investigated and compared. Hyperspectral mapping on the ACG was performed to obtain simultaneously TPPL and SHG maps. This ensures a fair comparison since the two nonlinear signals were generated on the same structure with the same excitation condition. The spatial distribution of the two nonlinear signals is analyzed to show the antenna role of the plasmonic grating. The difference in the spatial distribution is also discussed.

The ACG is designed in a way that the trajectory of the $n^{th}$ circular groove can be described as[12, 31]

$$(x - nd)^2 + y^2 = (n \Delta r)^2, \quad (1)$$

where $d$ is the displacement of the center of the circular grooves and $\Delta r$ is the increment in radius. The grating periodicity ($P$) at a certain in-plane azimuthal angle ($\varphi$) is given by,

$$P_{\Delta r, d}(\varphi) = \left| d\cos\varphi \sqrt{(d^2 \cos 2\varphi + 2\Delta r^2 - d^2)/2} \right|, \quad (2)$$

As the in-plane azimuthal angle increases from 0° to 180° the periodicity of ACG varies continuously from $\Delta r+d$ to $\Delta r-d$, as illustrated in Figure 2a. By choosing suitable $\Delta r$ and $d$, the range of grating periodicities can be easily designed for any desired application.[32, 33] With the azimuthal angle-dependent periodicity and the photon-plasmon phase-matching condition, an azimuthal angle-dependent resonance wavelength ($\lambda_0$) can be determined by,

$$\lambda_0 = \frac{|d\cos\varphi \pm \sqrt{(d^2 \cos 2\varphi + 2\Delta r^2 - d^2)/2}|}{m} \left( \sqrt{\frac{\varepsilon_m \cdot n_d^2}{\varepsilon_m + n_d^2}} - n_d \sin\theta \right), \quad (3)$$

where $m$ is the order of resonance, $\theta$ is the angle of incidence, $\varepsilon_m$ is the permittivity of the metal, and $n_d$ is the index of the dielectric surrounding. The designed ACG was fabricated by using a gallium focused-ion beam (FEI Helios NanoLab 600i) by sputtering selective areas of chemically synthesized gold flakes. The thickness of the flakes is estimated to be ~200 nm. The target milling depth is 150 nm and the width of the grooves is around 50 nm. The ACG used for the investigation of TPPL and SHG consists of twenty rings starting from the smallest ring with a zero radius. Outside of the grating area of the ACG, two pairs of horizontal and vertical slits are fabricated for position alignment. Figure 2a shows the schematic design of a typical ACG and the SEM image of one of the ACG used for TPPL and SHG investigation in this work.

For the investigation of SHG and TPPL generation from the plasmonic ACG, the wavelength-tunable (990 – 1100 nm) signal beam of an optical parametric oscillator (APE) pumped by a Ti:Sapphire oscillator (Mira HP, Coherent) has been used. The laser emits pulses of ~150 fs duration, an average power of ~200 mW, and a 76-MHz repetition rate for fundamental excitation. Details of the experimental setup are provided in the Supporting Information (Figure S1). SHG and TPPL were performed on the same ACG, and the results were acquired simultaneously. The laser beam was guided through a quarter waveplate (Eksma Optics, Crystal Quartz+MgF$_2$ achromat quarter waveplate, retardation L/4@950-1300nm) to obtain circularly polarized light before being focused onto the ACG through a microscope objective with a numerical aperture (NA) of 0.5 (Olympus, UPLFLN20X). Since the upper half and the lower half of an ACG are mirrored, we will only discuss the results from the upper half of the ACG. Sometimes, slight differences are observed between the two halves due to the imperfect circular polarization of the excitation. The mirror symmetry of the images from the two halves of the ACG thus serves as an indicator for optical setup optimization. Slight asymmetry due to elliptical polarization of the illumination can be corrected with simple post-data processing.[33]

The ACG was mounted on a three-axis piezo nano-positioning stage (Nano-PDQ series, Mad City Labs), allowing a raster scan over the grating structure. The generated SHG and TPPL spectra were then collected by the same objective and aligned through two short-pass filters (Semrock, Brightline multiphoton 770SP) to spectrally filter out the fundamental excitation beam before being focused by an achromatic lens to the entrance slit of the spectrometer (Andor, Kymera 193i) equipped with a CCD camera (Newton920, DU920P-BEX2-DD). With the position-scanned spectra acquisition, hyperspectral images were obtained for spatially and spectrally resolved investigation of the SHG and TPPL on the plasmonic ACG. As an example, the mean spectrum of simultaneously measured second harmonic and TPPL signal from ACG at an azimuthal angle $\varphi = 45°$ is shown in Figure 2b. As can be seen, the SHG peak is intensive and sharp while the TPPL is a broadband background decaying from the excitation wavelength towards a short wavelength regime.

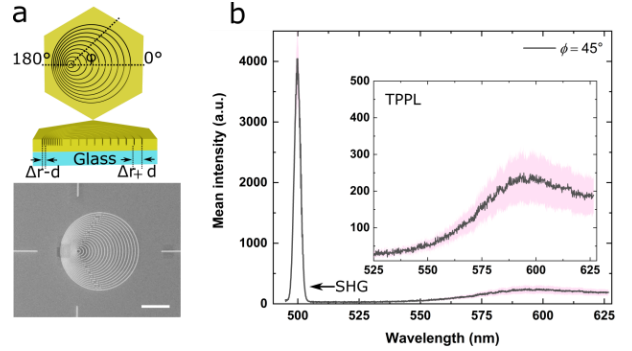

**Figure 2.** (a) The upper panel shows the design of ACG in the top view with the tilted view of the cross-section. The lower panel shows one of the SEM images of the gold ACG structures used in this work with $\Delta r$ = 700 nm, and d = 400 nm. Scale bar = 10 μm. (b) Mean spectrum showing simultaneously the SH peak and TPPL signal from ACG at an azimuthal angle of 45°. The inset shows the TPPL part of the spectrum. The error bar represents the standard deviation of the mean spectrum obtained from all points along the 45° in-plane azimuthal angle within the area of the ACG.

**Results and discussion**

The azimuthal angle range of the grating (the sector area) suitable for far-to-near field coupling can be calculated analytically using Eq. 3 with the knowledge of the objective NA and the excitation wavelength. As for the out-coupling of the generated nonlinear signal, a similar analytical calculation can be performed to obtain the sector areas suitable for emission out-coupling at the selected wavelength.

It is worth noting that Eq. 3 solely describes the grating-assisted photon-to-plasmon momentum matching. Therefore, it only predicts the allowed angle ranges fulfilling the momentum matching condition without providing any information on the field intensity because the latter also depends on the impinging laser conditions. Since our excitation beams have a Gaussian beam profile, most of the excitation power passes through the central area of the objective, corresponding to an effectively smaller NA than the specification of the objective. Therefore, most of the excitation power is delivered to the ACG within a small incident angle range. As a result, within the allowed in-coupling sector range for the excitation beam, the incident power is inhomogeneous with a gradient from the largest power on the right boundary at $\varphi$ = 35° to the weakest power on the left boundary at $\varphi$ = 78° (Figure 3a). The inhomogeneous intensity distribution is indicated in the analytically calculated sector area with gradient colors. The full width at half maximum (FWHM) of the Gaussian intensity profile corresponds to an azimuthal angle $\varphi$ = 62°. This gradient excitation power distribution needs to be taken into consideration when analyzing the spatial distribution of the enhanced nonlinear signals on an ACG.

We first focus on TPPL. Figure 3a shows the calculated sector areas suitable for the in-coupling of an excitation wavelength at $\lambda_{ex}$ = 1000 nm via an m = -1 grating order. The best out-coupling sector areas for TPPL emission at 595 nm via an m = -1 and -2 grating order are marked in Figure 3b. The consequent overlapping areas of the grating sectors for excitation in-coupling and emission out-coupling are shown in Figure 3c. In the overlapping areas, both the excitation and TPPL emission are enhanced and the maximum TPPL signal at the selected wavelength is expected. The experimental results shown in Figure 3d agree well with the analytically predicted distribution. The strongest emission signal appears at the grating within azimuthal angle ranges of 35° to 65°. The nice agreement between the experimental results and the calculation confirms that the plasmonic ACG serves as an optical antenna to provide spatially and spectrally resolved near-to-far field coupling. A movie demonstrating the hyperspectral mapping of TPPL and SHG signal from ACG is provided in the Supporting Information.

Next, we focus on SHG. The sector areas for the in-coupling of the excitation are the same as that for TPPL (Figure 3a), and the emission is at the SHG wavelength. The calculated grating areas for SHG emission out-coupling via grating orders m = -1, -2, and -3 are depicted in Figure 3e with the consequent overlapping areas shown in Figure 3f and the experimental intensity distribution displayed in Figure 3g. A strong SHG signal is observed within the analytically calculated sector areas suitable for an excitation in-coupling at $\lambda_{ex}$ = 1000 nm via an m = -1 grating order and the emission out-coupling at $\lambda_{SHG}$ = 500 nm via an m = -3 order. Another analytically predicted enhancement sector area between 63°-78°, where the out-coupling of the emission at $\lambda_{SHG}$ is enhanced via an m = -2 grating mode, does not show a strong SHG signal in the experiment. The signal is not strong since the excitation beam has a Gaussian profile, and the power within the excitation sector (Figure 3a) decays to the FWHM of the Gaussian intensity profile at about $\varphi$ = 62°. Therefore, the sector area between 63°-78° is receiving relatively weak excitation power. Since SHG is quadratically dependent on the excitation power, the signal decays even faster than the excitation power gradient, leading to a rather dim SHG signal.

An intriguing difference between TPPL and SHG can be seen in the low azimuthal angle range between $\varphi$ = 0° and $\varphi$ = 35°. While TPPL is almost undetectable in this range, SHG shows detectable intensity like that from the unpatterned area. Within this angle range, the ACG does not provide the momentum needed for the photon-plasmon coupling at the excitation wavelength. Therefore, no excitation enhancement is expected for TPPL or SHG. This difference in the intensity of TPPL and SHG in this angle range stems from the different nonlinear mechanisms and selection rules responsible for TPPL and SHG. As discussed previously, TPPL from gold is a nonlinear process involving the "sequential" absorption of two photons. The first photon triggers an intraband transition of electrons in the sp-band and leaves an sp-hole with a lifetime of about one picosecond.[27] The intraband transition relies on the broadband momentum of the plasmonic hotspots.[17, 28, 34] Therefore, TPPL is undetectable

on the unpatterned ultrasmooth gold flake surface due to the absence of plasmonic hotspots.[34]

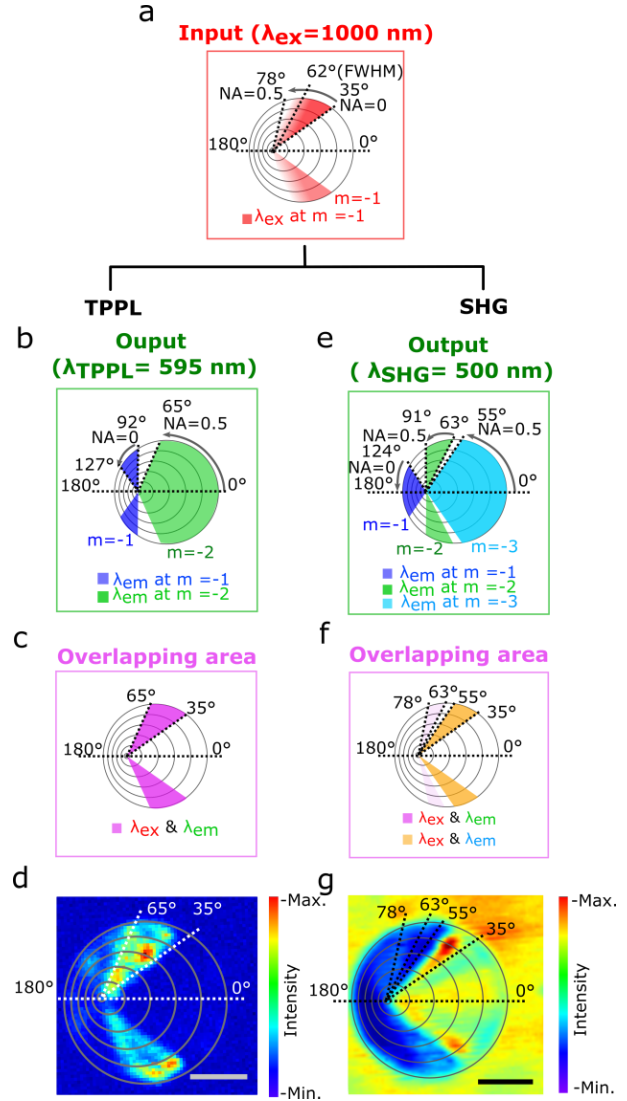

**Figure 3.** (a) Analytically predicted azimuthal distribution of gratings suitable for the in-coupling of the excitation wavelength ($\lambda_{ex}$) at 1000 nm. (b) Azimuthal distribution of the emitted TPPL signal ($\lambda_{TPPL}$) at 595 nm for different grating orders (m). (c) Overlapping areas of the sectors shown in (a) for TPPL excited at 1000 nm and emission at 595 nm (b). (d) Experimentally observed TPPL signal from the ACG. Scale bar = 10 μm. (e) Azimuthal distribution of emitted SH signal at 500 nm for different orders. (f) Overlapping area of the sectors shown in (a) and (e) for SHG excited at 1000 nm and emission at 500 nm, respectively. (g) Experimentally observed spatial distribution of the SH signal from ACG. Scale bar = 10 μm.

In contrast, SHG is a second-order nonlinear process that requires a broken symmetry of the oscillator or the modes responsible for the nonlinear signal generation.[35] In principle, gold is not suitable for SHG because it is a centrosymmetric material. However, symmetry breaking is enforced at the gold surface. Therefore, SHG from the unpatterned gold surface is allowed.[36] In the lower azimuthal angles range (particularly between $\varphi$ = 0° and $\varphi$ = 35°), the grating periodicity (1100 nm - 989 nm) is comparable to or larger than the size of the laser focal spot (diameter ~1000 nm). This means that the excitation beam does not see the grating but a flat gold surface. As a result, the efficiency of TPPL and SHG in this regime is very similar to that in the unpatterned gold surface area. This can be clearly seen in Figures 3d and 3g, where the signal from the low angle range is like that of the unpatterned surface. This difference reveals the different fundamental mechanisms responsible for the TPPL and SHG processes.

In summary, we investigated the enhancement effect of a plasmonic ACG in surface-enhanced TPPL and SHG. The ACG serves as a spatially and spectrally resolved antenna that mediates the far and near optical fields of multiple input and output beams at different frequencies. The intensity distribution maps the two nonlinear processes, which are investigated as a function of the varying periodicity of the ACG. TPPL and SHG signals were simultaneously generated by efficiently coupling both fundamental excitation and emission at a shorter wavelength on a single ACG nanostructure. This study demonstrates that resonant plasmonic gratings serve as optical antennas to mediate the far and near optical fields. The difference in the spatial distributions of SHG and TPPL is revealed and linked to the responsible mechanisms. This work provides valuable information for the design of effective plasmonic nanostructures for nonlinear optical processes.

ASSOCIATED CONTENT

SUPPORTING INFORMATION

Experimental setup, and video of TPPL and SHG intensity mapping.

AUTHOR INFORMATION

Corresponding Author

**Jürgen Popp** Leibniz Institute of Photonic Technology, 07745 Jena, Germany; Institute of Physical Chemistry and Abbe Center of Photonics, Friedrich-Schiller-Universität Jena, D07743 Jena, Germany;
Email: Juergen.popp@leibniz-ipht.de

**Jer-Shing Huang** – Leibniz Institute of Photonic Technology, 07745 Jena, Germany; Institute of Physical Chemistry and Abbe Center of Photonics, Friedrich-Schiller-Universität Jena, D-07743 Jena, Germany; Research Center for Applied Sciences, Academia Sinica, Taipei 11529, Taiwan; Department of Electrophysics, National Yang Ming Chiao Tung University, Hsinchu 30010, Taiwan;
Email: jer-shing.huang@leibniz-ipht.de

ACKNOWLEDGMENT

The financial support from DFG via CRC 1375 NOA (Subprojects C1 and C5) and the European Union via Horizon 2020 research and innovation programme (Grant agreement No 101016923) is gratefully acknowledged. The authors thank Yi-Ju. Chen and Oliver Rüger for the support in data analysis and sample fabrication.

ABBREVIATIONS

ACG, azimuthally chirped grating; SHG, second harmonic generation; TPPL, two-photon photoluminescence.

TOC figure

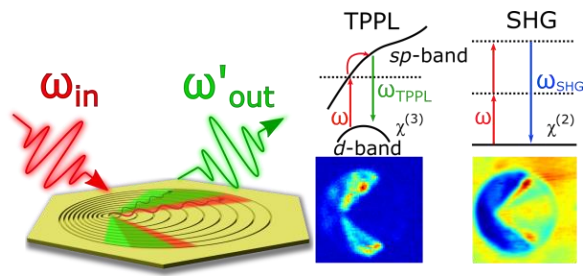

# Supporting information

# Nonlinear optical signal generation mediated by a plasmonic azimuthally chirped grating

Parijat Barman,[†,‡] Abhik Chakraborty,[†,‡] Denis Akimov,[†,‡] Ankit Kumar Singh,[‡] Tobias Meyer-Zedler,[†,‡] Xiaofei Wu,[‡] Carsten Ronning,[⊥] Michael Schmitt,[†,‡] Jürgen Popp,[†,‡,*] Jer-Shing Huang[†,‡,§,#,*]

[†]Institute of Physical Chemistry and Abbe Center of Photonics, Friedrich-Schiller-Universität Jena, Helmholtzweg 4, D-07743 Jena, Germany

[‡]Leibniz Institute of Photonic Technology, Albert-Einstein Str. 9, 07745 Jena, Germany

[⊥]Institut für Festkörperphysik, Friedrich-Schiller-Universität Jena, Max-Wien-Platz 1, 07743 Jena, Germany

[§]Research Center for Applied Sciences, Academia Sinica, 128 Sec. 2, Academia Road, Nankang District, Taipei 11529, Tai-wan

[#]Department of Electrophysics, National Yang Ming Chiao Tung University, Hsinchu 30010, Taiwan

*Correspondence should be addressed to:
J.P. (Juergen.popp@leibniz-ipht.de) and J.-S. H. (jer-shing.huang@leibniz-ipht.de)

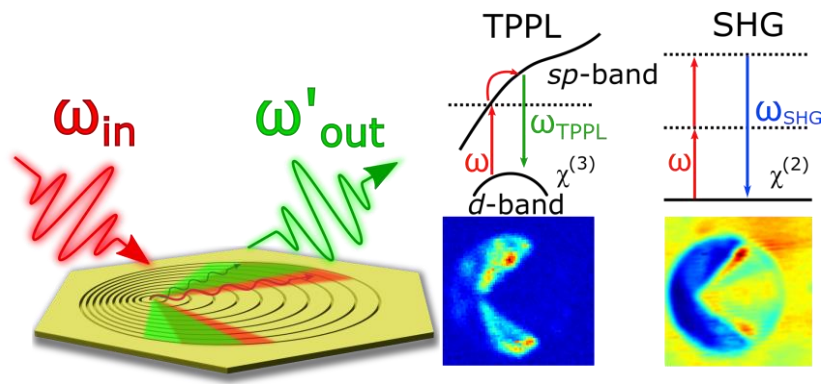

**Content:**

I.  Experimental setup (Figure S1)

II. Video of TPPL and SHG signal of ACG

## I. Experimental setup

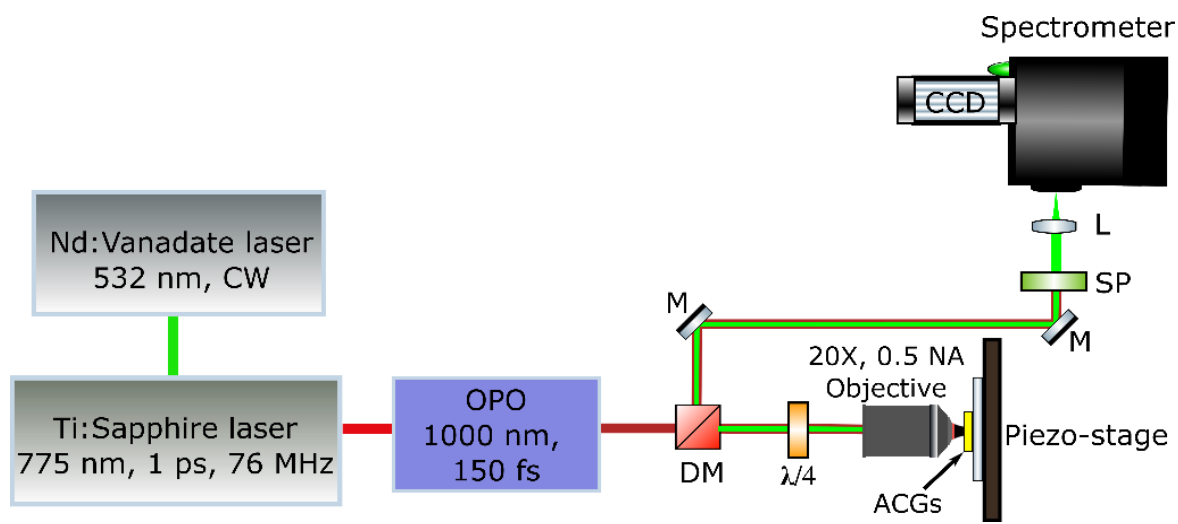

**Figure S1.** Setup for the TPPL and SHG experiments. A Ti: Sapphire laser (pumped by Nd: Vanadate laser) is used as pump source for an optical parametric oscillator (OPO) to generate 1000 nm (tunable) ~150 fs pulse as fundamental excitation. DM: dichroic mirror, M: mirror, λ/4: quarter wave-plate, SP: short-pass filter, L: lens

## II. Video of TPPL and SHG signal of ACG

''ACG_TPPL_SHG.avi"

The movie shows the hyperspectral mapping of TPPL and SHG signal distribution on ACG.